# Extending the Web3D: Design of Conventional GUI Libraries in X3D

Ivan Sopin
Computer Science
Armstrong Atlantic State University
11935 Abercorn Street
Savannah, GA 31419
+1 (912) 401-1407

ivansopin@gmail.com

Felix G. Hamza-Lup
Computer Science
Armstrong Atlantic State University
11935 Abercorn Street
Savannah, GA 31419
+1 (912) 344-2680

felix.hamza-lup@armstrong.edu

## ABSTRACT
Extensible 3D (X3D) modeling language is one of the leading Web3D technologies. Despite the rich functionality, the language does not currently provide tools for rapid development of conventional graphical user interfaces (GUIs). Every X3D author is responsible for building—from primitives—a purpose-specific set of required interface components, often for a single use.

We address the challenge of creating consistent, efficient, interactive, and visually appealing GUIs by proposing the X3D User Interface (X3DUI) library. This library includes a wide range of cross-compatible X3D widgets, equipped with configurable appearance and behavior. With this library, we attempt to standardize the GUI construction across various X3D-driven projects, and improve the reusability, compatibility, adaptability, readability, and flexibility of many existing applications.

## Categories and Subject Descriptors
D.1.7 [**Programming Techniques**]: Visual Programming; D.2.2 [**Software Engineering**]: Design Tools and Techniques – *modules and interfaces*; D.2.6 [**Software Engineering**]: Programming Environments – *graphical environments.*

## 1. INTRODUCTION
With 3D graphics firmly entering the domain of the Internet, a new niche of virtual visualization—called Web3D—has formed. One of the leading technologies united in the realm of Web3D is the Extensible 3D (X3D) modeling language. Due to the immense graphical and scripting capabilities, X3D has become a mature graphics standard with wide recognition among professional organizations, researchers, 3D designers, and Web3D enthusiasts around the globe. Yet one important feature that the language still lacks is a toolset for creating conventional user interfaces (UIs). Typically, to provide a UI for each new application, the X3D author has to design an entirely different set of interface components. Most of these implementations are very limited in functionality and only serve their ad-hoc purpose.

We address the challenge of creating consistent, efficient, easily controllable, and visually appealing interfaces by proposing the Extensible 3D User Interface (X3DUI) library. This library is a wide range of cross-compatible X3D widgets with configurable appearance and behavior. With X3DUI, we attempt to standardize the UI construction across various X3D developments and improve the reusability, compatibility, adaptability, readability, and flexibility of many existing applications. The library is composed of traditional Microsoft-Windows-like UI elements, whose configuration parameters in general correspond to analogous realizations in Java or Visual C++.

We further argue that humans are technologically more accustomed to planar interface layouts, and thus the management of virtual 3D content in X3D can be effectively performed through 2D or 2.5D (2D with seeming depth) UIs, rendered using the heads-up-display (HUD) technique. Another reason for reduced dimensionality of the UIs is that the truly 3D interfaces are difficult to operate and deliver via inherently 2D visualization systems, such as computer monitors and overhead projectors.

The rest of this paper is organized as follows. The second section presents current research and development efforts in the field of graphical UIs (GUI) design for X3D-based visualization systems; the common usability- and interactivity-related issues are analyzed. In section 3 we examine various presentation-specific aspects of X3DUI and illustrate several usage scenarios. The organization and implementation of the library components are elaborated in section 4. We conclude in section 5 with a brief summary and considerations for future work.

## 2. RELATED WORK
### 2.1 Interface Classification
In this section we provide an overview of existing GUI solutions for interactive X3D simulations. Some developers suggest X3D-only implementations, with interface components constructed using the native language definitions—similarly to the library proposed here. Others employ X3D content as a part of composite multimedia environments, backed up with either conventional technologies, such as HyperText Markup Language (HTML) and JavaScript, or entire proprietary frameworks and application programming interfaces (APIs). We discuss the flaws and merits of each approach and analyze some of the common issues.

## 2.2 X3D-Based Interfaces

The most straightforward method to create a GUI in X3D is to interconnect suitable geometric nodes via natively supported scripting logic. It is, however, difficult to achieve an aesthetically pleasant, functionally rich, and programmatically convenient architecture by dealing with geometric primitives and low-level spatial transformations. For this reason, the overwhelming majority of existing X3D-based GUIs merely feature a few basic components to support the minimum required functionality. In what follows we present several more advanced examples.

In 2006, two X3D-driven Web-based simulation tools for radiation therapy planning procedures were built as a part of 3D Radiation Therapy Treatment (3DRTT) project [1]. The simulators incorporate a set of floating HUD-menus, containing toggle buttons, sliders, and scrolls for manipulating the virtual linear accelerator hardware (Figure 1). The design of the interface follows the traditional approach, described earlier, and only targets one ad-hoc purpose. Despite the simplicity and intuitiveness, this GUI takes up a lot of screen real estate and renders poorly on very high and low zoom levels as well as in stereo mode (because of the false focal distance, as discussed further in section 3.2). Additionally, the menus are prone to visual collisions and can be accidentally moved off the screen.

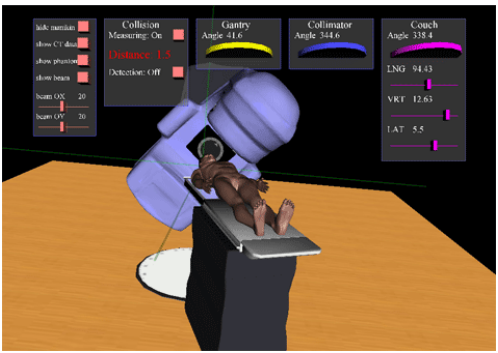

**Figure 1. Linear accelerator simulator with X3D-based GUI.**

A more systematic methodology is practiced in the CONTIGRA architecture [2], where Extensible Markup Language (XML) is used to define GUI schemes for the resulting X3D world. The architecture comprises three major levels: `SceneGraph`, `SceneComponent`, and `Scene`. At the `SceneGraph` level, X3D entities are used to define the geometric components; special grammars are introduced to program the sets of additional nodes for extended behavioral and audio functionality. `SceneComponent` serves as the markup language of the CONTIGRA architecture and is used to provide interface declarations as well as detailed widget configurations. The `Scene` level governs the integration procedure and yields executable X3D code. This realization demonstrates powerful abstraction techniques that eliminate bindings to different implementation frameworks, while allowing writing high-level format-independent code. Nonetheless, it is still the developer's responsibility to create the ultimate building blocks of the GUI, most likely as complementary `SceneGraph` nodes. Developing such an extension using CONRIGRA architecture could be more complicated than constructing a GUI library in X3D directly. A similar approach, but with the emphasis on XML and Extensible Stylesheet Language Transformations (XSLT), is presented in [3], where a sequence of inter-format translations is performed to analyze, interpret, and merge the model and the scene files into an X3D scene.

Another example of multilayer architecture is explained in [4]. It employs the UsiXML [5] user interface definition language (UIDL) to describe the scene, which is subsequently converted via VUIToolkit into X3D code for rendering. Fast deployment, however, does not provide high universality and usability of the generated interface. Applicable for individual models, this approach lacks the overall design completeness and integrity.

The common problems of the existing X3D-based GUI implementations include the inability to cover dynamic changes of the interface structure and configuration; necessity to compile each interface individually; questionable rendering quality; and unnatural fusion with the 3D portion of the scene. Utilization of special development tools further reduces the fitness of such solutions for wide use among X3D content creators.

Next we will review a series of HTML-based interfaces that overcome many of the presentation- and usability-related shortcomings of X3D-based GUIs.

## 2.3 HTML-Based Interfaces

To visualize and interact with an X3D world, the user needs special player software. There are currently over a dozen X3D players, with different distribution licenses and levels of X3D component support. Some of these players can be installed both as a stand-alone application and a Web browser plug-in, the latter enabling X3D scenery to blend smoothly into the context of a Web page. With such functionality in place, the interactive functions of the interface can be effectively delegated to various multimedia ingredients of the page, including the X3D world, HTML controls, JavaScript scenarios, and so on. Most X3D plug-in manufacturers provide a simple API for runtime access to the geometrical and scripting nodes of the scene via JavaScript code run in the scope of the host page. The backward capability to invoke predefined JavaScript methods from within the X3D environment is often provided as well.

The potential of multimodal HTML-based interfaces for online X3D simulations has been explored in a number of projects developed in the NEWS laboratory [6]. Early versions of one of them, 3DRTT, were discussed in the previous section. The current generation of simulators presented in the 3DRTT project contains a more sophisticated GUI that relies on HTML and Cascading Style Sheets (CSS) for visual representation, and JavaScript for functionality. Control over the virtual environment is ensured by frequent invocations of internal X3D procedures, conducted from the Web page's JavaScript code.

The nervous system simulator from the NeuroPathways project (coordinated by the NEWS as well) also features an HTML-based interface for controlling the X3D scene. The simulator additionally uses Asynchronous JavaScript and XML (AJAX) technology to dynamically send HTTP requests to the server; the server analyzes these requests, queries or updates the database, and replies with an appropriate response. Such architecture is known as AJAX3D, and is designed to combine the benefits of real-time 3D with the power of Web-enabled interfaces. AJAX3D enables the developer to dynamically manage X3D worlds with JavaScript via the Scene Access Interface (SAI).

The Ludos Top project [7] from the Federal University of Uberlândia, Brazil, is another showcase of the AJAX3D paradigm. More specifically, the project demonstrates the feasibility of designing real-time multiplayer 3D online games by expanding the existing Web deployment architectures with the X3D module. The AJAX3D schema operating on top of the Linux/Apache/MySQL/Python software stack, with additional template extensions, also proved effective in the WebScylla [8] project. In this case, an eye-catching HTML-based GUI is combined with realistic X3D animations to allow the user to interactively visualize the colonization of an artificial reef.

Ultimately, the cohort of present-day HTML-based GUIs for X3D simulation systems reveals certain strong sides of Web3D: interactivity, accessibility, and compatibility with other Web technologies. However, the HTML-based-GUI metaphor entails serious limitations on how the X3D content is presented: The scene can only be interacted with inside the browser. The GUI relies on JavaScript communication, often very intensive, which results in high computation costs and produces delays and jitter in the final visualization. HTML includes only a subset of interface components commonly used in operating systems (OSs); for instance, windows, sliders, and tab panels are not normally supported in a Web page. GUIs cannot be easily rendered in stereo mode and are somewhat tied to the screen size and resolution. Elements of the interface usually cannot render above the 3D scene and thus reserve substantial area of the screen. Transparency and visibility are difficult to implement without disrupting the consistency of the GUI. Lastly, sophisticated GUIs with intricate object positioning might be browser-dependent.

A notable effort to consolidate X3D and HTML technologies in the next-generation Web-page-language standard is made in the X3DOM project [9]. The proposed framework is considered for inclusion into the HTML5 specification as a way to support declarative 3D content—defined by X3D nodes—natively in the document object model (DOM). This improvement would eliminate the need of a browser plug-in, and enable the 3D scene management via JavaScript operations on DOM tree rather than on SAI (which could also be plug-in-specific).

## 2.4 Other Approaches
While most commonly GUIs for X3D visualizations are implemented internally or in conjunction with HTML, several research teams attempted to invent an entire new UIDL to universalize the interface generation. For example, GUIML is proposed as an interface markup language for Web3D [10]. The language is described in XML and is comprised of multiple interface components and callback interaction mappings. The authors argue that GUIML has the potential to become a standardized method of GUI abstraction, X3D being a particularly promising 3D interface-instantiation environment. However, the architecture requires a special interpreter to carry out the conversion and is better suited for general-case projects.

We have shown various trends and techniques for creating GUIs in Web3D applications powered by X3D. We have also provided a simple interface classification to point out the virtues and weaknesses of major design approaches; a more elaborate evaluation framework for Web-based visualizations is presented in [11]. In the remainder of this paper we expound on how our solution targets the problems described, rationalize the utilized design patterns, and work out the details of the implementation.

## 3. DESIGN
### 3.1 Dimensionality
Numerous technical innovations have entered our lives over the last few decades. The means and media of social communication—and information exchange as a whole—have drastically changed. Information is accessed, perceived, and stored in various new configurations and formats. What has remained unchanged for the most part is the 2D nature of content delivery. Maps, newspapers, billboards, Web sites, and TV broadcasting make good examples. Despite living in a 3D space, we are traditionally used to 2D arrangement of information, which provides for effective distribution and interpretation. The intrinsic storage and presentation complexity of physical 3D displays also contribute to this condition. As a result, the vast majority of human-designed interfaces are semantically 2D; that is, having three physical dimensions, including depth, such interfaces are logically confined to their planar equivalents. Some of the customary widgets incorporated in these interfaces include buttons, switchers, sliders, and so on.

The same phenomenon extends to the realm of virtual 3D: Whereas the volumetric representation of the scenery is important; 2D or 2.5D (2D with simulated depth) implementations are usually more advisable for GUIs [12]. The symbiosis of two- and three-dimensional graphics has proved to be an effective solution for many computer applications, such as video games, modeling software, and various medical, engineering, aeronautic, architectural, and educational simulations.

Practical visualization systems should be able to interact with the user by accepting some sort of human control and generating the proper responses. Generally, the more intuitive and submissive is the UI, the better is the overall operability of the system. In [13] it is shown that the precision of user manipulations could be enhanced by applying simulated surface constraints to the 3D interface of a virtual environment, which once more proves the better fitness of 2D and 2.5D realizations for GUI design.

This principle lays in the foundation of X3DUI. We have attempted to build a library that preserves the power of X3D to deliver rich volumetric content, while supplying the user with convenient, effective, and—more importantly—conventional tools to control it. Next, we explain the visualization characteristics of X3D and outline how those are applied in X3DUI.

### 3.2 Presentation Techniques
Despite the seeming easiness, in practice the systematic incorporation of 2D and 3D graphics within one scene is a non-trivial task, especially when dealing with such a crucial ingredient of the visualization as the UI. Because the interface dictates strict accessibility requirements, it is normally visualized in the HUD-layer, in front of everything else. Numerous questions arise regarding the applicability of this technique: Will the UI permanently eclipse the objects in the background? How will avatar orientation changes be treated? How will the interface react to zooming? Could the background scenery partially penetrate the interface controls? These and other questions are addressed in the logic of the X3DUI architecture, as explained next.

To accomplish the HUD-like behavior in X3D, a method of routing a `ProximitySensor` to a `Transform` node containing the targeted geometry has been adopted: As the user navigates through space, the `ProximitySensor` detects the viewpoint position and orientation changes, and, using the `ROUTE` construct, updates the `Transform`'s translation and rotation fields. The result of this coordination is that visually every child node contained in the `Transform` appears to remain unaffected by the perspective displacement.

Unfortunately, the approach described has several issues, which manifest themselves differently in different X3D players. For instance, when the scene is zoomed in too close or zoomed out too far, the geometry of the HUD-layer might flicker in some players; in others, it will "shake" on any orientation change. Additionally, HUD-objects are penetrated by movable objects that are very near to the viewpoint. Finally, with increasing interest in simulated volumetric visualization, the ability to render a virtual scene in stereo mode becomes very relevant. In case of HUD-based layers the stereo mode produces widely diverged separation for right- and left-eye views, resulting in a bifocal decomposition of the scene. This phenomenon is stimulated by the false focal distance of the HUD layer, which does not obey to the spatial transformations in the scope of global coordinate system.

As an alternative to the HUD technique, X3DUI uses the `Layer3D` node, provided by BitManagement via a prototype, and hence easily transportable to other players. `Layer3D` allocates a transparent rectangular area of the screen to render an autonomous scene by overlaying it on top of the host scene. This node enables smooth layering and is free of penetration- and stereo-rendering-related problems.

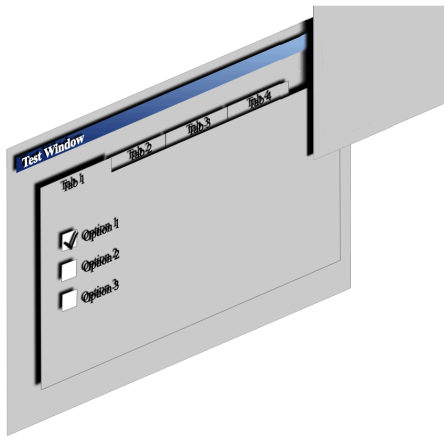

**Figure 2. GUI depth-separation example.**

A bigger challenge than merging planar and volumetric geometry in 3D visualization systems, and X3D, in particular, is consistent management of 2D layers in the shared z-plane. Naturally, GUI components placed within one container are rendered at an equal distance to the viewpoint, and nothing prevents their surfaces from interpenetration. This mixture is easily avoided by slightly dispersing the layers along the z-axis. However, in large sets of GUI components, even an insignificant dispersion increment might contribute to an oversized separation range (Figure 2), which is very apparent at side-by-side comparison of the affected nodes. If—based on value, order of use, or relationship—certain pieces of the interface have to be readjusted in the depth stack, noticeable visual permutations are generated. A more appealing, yet even more intricate solution is to instruct the renderer to display in a specific order the 2D items that coincide in z-depth.

We program this behavior in X3DUI using `OrderedGroup` extension node from BitManagement. The node constitutes a simple container, with rendering priority of its children specified in the array-type `order` parameter. Because sensor nodes in X3D do not take into account special rendering effects, mere adoption of `OrderedGroup` is not sufficient to disambiguate the scopes of overlapping sensors. This is why X3DUI also employs the depth-separation technique in cases when accessibility is more important than perfect appearance.

The next essential aspect of utilizing GUI overlays is the occlusion of non-GUI scenery. A possible method to improve the usability is to introduce an appropriate transparency level to the interface components and make them "hideable" or closeable. These effects are readily obtainable in X3D with the aid of `Material` and `Switch` nodes combined with simple scripting. Extra degrees of freedom may be provided by making the GUI controls resizable and draggable across—but not beyond—the screen. The cost of such enhanced interactivity is the higher complexity of implementation. To facilitate the designer's work, X3DUI library incorporates multiple techniques for efficient use of screen real estate: resizable, minimizable, and closeable floating windows; support of transparency by all visual components; and control over the size of most objects.

One more crucial item in most virtual GUI designs is text. Not only are text-driven interfaces informative, but sometimes essential to the understanding of a control's function; even more so if no explicit link exists between the trigger and the event. Nonetheless, text rendering can be more elaborate than rendering of geometric primitives, the reason being a high polygonal count caused by tesselation. To "fill up" the character contours, a great number of varied-size polygons are clustered together in a mesh. While transformations are applied to the text object, every polygon in the mesh is updated with the new transformational matrix, resulting in the visual disruption among adjacent polygons within one character entity. By supplying the `USE_TEXTURE` flag in the style attribute of `FontStyle` node the Contact player can be instructed to render the associated message using a texture, and therefore avoid tesselation-related problems.

We have covered several issues of the "2D-in-3D" GUI metaphor that apply to the implementation of X3DUI. More details, regarding individual components of the library are discussed in section 4.

## 4. IMPLEMENTATION

### 4.1 Structure Overview

Currently, X3DUI includes twenty-seven prototypes classified into four categories: *system*, *visual*, *group*, and *layout*. The *system* category includes prototypes which organize the work of all widgets and are imperative to the functioning of the entire library.

All prototypes with a visual representation are collected in the *visual* category. The *group* category holds prototypes that manage the behavior of several nodes in one group. Lastly, the prototypes for laying out elements within a `Panel` or `Frame` container are combined in the *layout* group.

Due to obvious reasons, the most inherited prototype in the X3DUI library is `Rectangle`. Not only many ordinary GUI components have rectangular shape, but they are also better described in 2D space in terms of their width and height. This convention enables easier grouping and packing of interface nodes in higher-level node containers, such as `Panel` and `Frame`. In fact, the entire functioning of any layout manager rests upon requesting or determining the dimensions of an element before it can be properly positioned.

## 4.2 Core Components
### 4.2.1 Display
`Display` is the central prototype of X3DUI, for it manages the operation of the entire interface. Implemented as a singleton, in `children` attribute this prototype encloses an array of the `Frame`-type objects. All children, along with a `TaskBar` node, are settled within an instance of `Layer3D` node, in the body of the prototype; `Script`, `MouseSensor`, and several associated `ROUTE`s follow. The primary functions of `Display` are propagation of unique identifiers among GUI nodes participating in the scene graph; window-overlay management; focus management; disambiguation of overlapping touch sensors' scopes (via intercepting mouse-triggered events); and status synchronization between the windows and the task bar.

### 4.2.2 Settings
`Settings` prototype contains various configurations that define the overall "look-and-feel" of the GUI. A single instance of this prototype is distributed recursively among all visual nodes. The configuration fields are declared `initializeOnly` for encapsulation purposes, and therefore may not be altered after initialization. Setter-methods for `Settings`'s fields are built into `Display`.

Under the default configuration, the appearance of X3DUI should be satisfactory for most users, and will enable swift and well-coordinated interactions. If modifications are desired, the authors can fine-tune the settings and evaluate the results locally before publishing the product. Future contributors to X3DUI will also be able to stylize the library by composing customized themes applied to graphical components.

One attribute of `Settings` that deserves special attention is `DEBUG`, which controls X3DUI logging. Because error messages displayed by many X3D players, including BitManagement Contact, can be very scattered and non-explanatory, identifying the faulty element at the time of debugging becomes a particular burden for the developer. X3DUI integral logging clarifies the GUI initialization sequence and facilitates the search of problematic code snippets.

## 4.3 Basic Visual Components
### 4.3.1 Rectangle, Layer, and Plane
`Rectangle` is a container with a small set of basic parameters inherited by most higher-level prototypes. Visually, `Rectangle` represents a box of certain size, visibility, color, transparency, and border type; it can host any number of other visual objects. Normally, this prototype should not be instantiated directly, but could be legitimately used as an immediate child to `Frame`.

The primary function of `Layer` is to allow dragging UI objects across the screen. Despite the small field-set, this prototype carries out very critical tasks behind the scene. For example, `Layer` ensures that a child remains entirely in the view, remembers its coordinates, and updates its location when the X3D player window is resized. Thanks to `Layer`, windows can be moved, maximized, and normalized in X3DUI.

`Plane` is a descendant of `Layer`, and is conceived as a "floating" alternative to `Rectangle`; the method-sets of the two prototypes are identical, apart from the motion-related extension inherited from `Layer`. `Frame` is currently the sole descendant of `Plane`, part of the reason being that `Frame` is the only permissible root-level visual node type, and hence all its children are also movable with the parent.

### 4.3.2 Button and ToggleButton
As a member of X3DUI, `Button` is the most primitive interactive GUI component, and has only two event-like `outputOnly` functions: `isPressed` and `isClicked`. The first function is called when the button is either being pressed or released, and the second—only upon pressing and then releasing the button. Both functions generate boolean values.

The main distinction of `ToggleButton` from `Button` is the persistence of unpressed and pressed states between user interactions. In other words, once "pressed" by the user, a toggle button will remain pressed until the user outpresses it.

## 4.4 Conventional Visual Components
### 4.4.1 TextButton, TextToggleButton, and CheckBox
`TextButton` is the X3DUI's implementation of arguably the most traditional GUI control: a rectangular button containing a text label with the summary of the performed function. This prototype extends `Button` with text-related functionality and one additional graphical state in the animation stack.

Analogously to `TextButton` descending from `Button`, `TextToggleButton` is `ToggleButton`'s child. Because of the visual "sticking" in the pressed mode, this prototype has two additional animation states as compared to `TextButton`.

A graphically autonomous component, `CheckBox` basically matches the functionality of `TextToggleButton`. However, `CheckBox` does not generate mouse-triggered events per se, such as on pressing or releasing a mouse button; this prototype only notifies the listeners about changing its status from being checked to unchecked, and vice versa.

### 4.4.2 RadioButton and RadioButtonGroup
`RadioButton` is semantically very close to `CheckBox`, and even shares similar graphical states. However, `RadioButton` can be checked, but cannot be unchecked by clicking on it. The justification of such behavior is that radio button objects should normally appear in groups where only one out of several options can be picked at any time (ensured by `RadioButtonGroup`), versus the group of check boxes, where each item is independent from the rest.

### 4.4.3 ControlButton

Despite being a typical GUI item, `ControlButton` is accessory to the `Frame` prototype. There are four subtypes of a control button, each with a unique pictogram and a distinct purpose. These subtypes are internally represented with the following self-explanatory flags: `MINIMIZE`, `MAXIMIZE`, `NORMALIZE`, and `CLOSE`. Normally, up to three of these buttons—generally including the minimizing and closing buttons—are placed on the right of a window's header. Since the maximized and normalized states of the window are codependent, the corresponding control button variants should not be put together. The window will automatically update the buttons upon the change of its status.

### 4.4.4 Label

`Label` takes care of the in-scene text management, and supports various text-specific characteristics, such as justification, font size, and font style, as well as multiline layout. If the preferred width of the label is smaller than some of the lines, the text in those lines will be ellipsized to fit; if impossible, the width of the component will be enlarged to at least accommodate the ellipsis. In the event of vertical overflow, the height of the label is increased to the aggregate height of all lines of text.

### 4.4.5 TextField

When the virtual environment requests a textual response from the user, a keyboard might be better suited for interacting with the UI than a mouse. In our library, keyboard input is provided via the `TextField` prototype. This prototype generates a rectangular field for viewing and editing an optionally predefined string of characters. `TextField` recognizes typed characters in lower and upper cases; allows deletion by 'Backspace'; finishes editing by 'Enter' and 'Escape' keys; and supports basic navigation using 'Home', 'End', and the arrow buttons on the keyboard.

While `TextField` is activated, a blinking cursor indicates the position of the caret. If the width of the field becomes insufficient to display the entire message, only the work section of the string—determined by the location of the cursor—is shown. It is also possible to specify the maximum length of the message by setting the `maxLength` field.

### 4.4.6 ComboBox

The `ComboBox` prototype is our implementation of a common drop-down menu control combined with a text field for editing. A peculiarity of `ComboBox` is that the drop-down list is drawn over other GUI elements. Not only should the list completely cover the graphics behind it, but should also remain operable.

### 4.4.7 HorizontalSlider and HorizontalRunner

`HorizontalSlider` is one of the more sophisticated visual prototypes in the X3DUI library. Besides the base appearance settings, the configuration of `HorizontalSlider` is composed of such parameters as minimum, maximum, and selected value; number of mark intervals and their exposure; discrete or continuous selection; and text in the left and right captions. This prototype can operate on negative numbers, work with ascending and descending intervals, and is capable of dynamically detecting and correcting invalid numeric bounds as well as out-of-range selection values. Similar functionality will be programmed into the `VerticalSlider`, which is not yet a part of X3DUI.

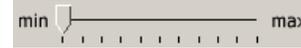

**Figure 3. `HorizontalSlider` object.**

Figure 3 demonstrates a horizontal slider control instance. `HorizontalSlider` takes advantage of the supplementary `HorizontalRunner` prototype, which defines the draggable pentagon-shaped runner. Use of highly tailored subcomponents helps us decompose the overall architecture and apply the divide-and-conquer strategy to our implementation.

## 4.5 Container Components

### 4.5.1 Panel

As in physical interfaces, individual components of virtual GUIs are better pronounced, perceived, and handled when they are arranged into logically cohesive groups. Moreover, coherent organization generally improves the mnemonics of a design. A computer keyboard is a perfect example: buttons with homogenous functions form a number of spatially disjoined blocks; disposition of these blocks is governed by the principles of memorability, usability, and ergonomics. We believe that similar reasoning ought to be embraced in the implementation of visual interface components in X3D.

Panel is the most basic and arguably the most universal grouping container in our library. The `Panel` prototype can nest multiple X3DUI nodes in a rectangular area with an optional border of lowered, raised, or edging style. The children are positioned according to the specified layout, defaulted to `FlowLayout`. If the initial area of the panel is too small to fit all elements, it is adjusted to the minimum qualified size.

### 4.5.2 TabPanel

Economy of space and bent for compactness have driven the developers to create a GUI control that would allow to both cram several sets of smaller widgets into one confined area and also provide unhampered access to them. This is how the tab panel came into existence. At any given time, the tab panel displays the content of the tab that was activated most recently. Selecting a tab is accomplished by clicking on the corresponding labeled header.

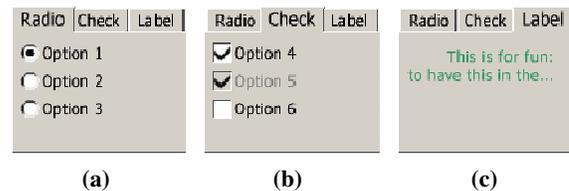

**(a)**      **(b)**      **(c)**

**Figure 4. `TabPanel` with the first (a), second (b), and third (c) tab activated.**

This mode of operation is threaded into the logic of `TabPanel` prototype. Figure 4 shows how a `TabPanel` object graphically adapts to the alternation of active tabs.

### 4.5.3 Frame

`Frame` derives its name from analogous class in Java programming language, and conceptually implements a window. A `Frame` instance can contain a header with the title and a set of control buttons, populated in accordance with the chosen minimization, maximization, and closing flags. A window may be also defined as floating or docked, and also resizable or static. Identically to the `Panel` prototype, `Frame` supports layouts.

The `Frame` widget resides in either active or inactive state (seen in Figure 5); activation is the result of any interaction with the widget. However, only a single window can be active simultaneously in an X3DUI-powered virtual desktop. Therefore, by actuating an inactive window, the user causes all other windows to become inactive, as follows from Figure 6.

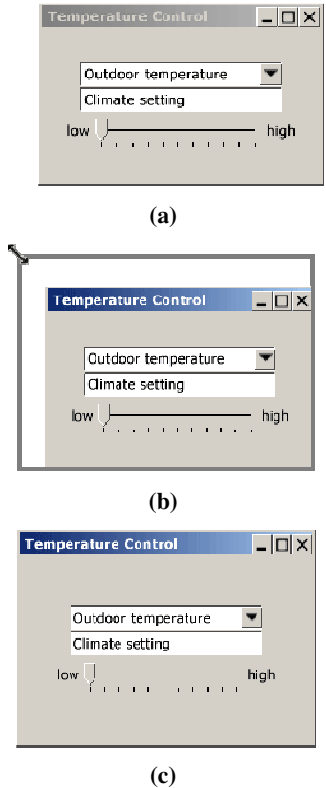

**(a)**

**(b)**

**(c)**

**Figure 5. `Frame` object in inactive (a), resizing (b), and resized (c) states.**

Resizing mechanisms programmed into the `Frame` prototype merit special consideration as an unprecedented implementation of that feature in X3D. As a rule, interface components created in X3D lose in usability and operability to their OS-specific counterparts; so, Web3D authors generally avoid complicating the interface design with additional graphical and behavioral features. In case of X3DUI, the experience of resizing a window demonstrates a high level of intuitiveness and robustness, as presented in Figure 5.

*4.5.4 TaskBar*
Task bar is integral component of many virtual desktop environments. Improvement of space efficiency is one purpose that it has in common with a tab panel; only instead of tabs the task bar controls windows. Respectively, in X3DUI the `TaskBar` prototype is designed to manage `Frame` prototype instances. Each opened `Frame` object is represented with a self-titled `TextToggleButton` control positioned within the `TaskBar`, in the order of creation.

Figure 6 serves as an illustrative example of `TaskBar`-enabled scene: Three windows, of which one is inactive and one is minimized, are duplicated with a set of desk-bands at the bottom of the screen. The middle desk-band, corresponding to the currently activated window, is pressed down.

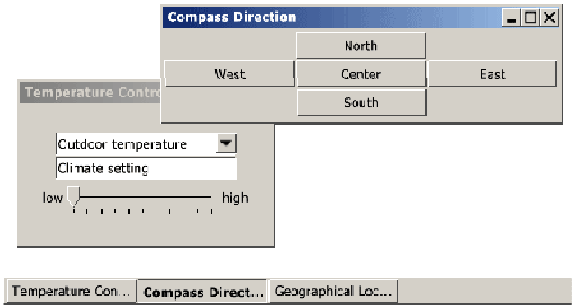

**Figure 6. A virtual desktop using the `TaskBar` prototype.**

Depending on the present status of a `Frame` object, clicking the respective desk-band can have different effects. If the `Frame` is minimized or just inactive, it will become active and will be brought to the front. In case of the already active `Frame`, the window will be minimized.

## 4.6 Layouts
*4.6.1 LayoutManager*
With the prevalence of object-oriented-programming-derived hierarchical visualization and success of programmatic layout managers, absolute positioning becomes more and more extinct in the modern GUIs. Exploitation of absolute coordinates results in static interfaces that disregard the client's preferences and often require total recasting upon insignificant rearrangements. A properly chosen layout, in contrast, can reduce the designer's job of organizing and maintaining the GUI to a minimum.

X3DUI is stocked with four popular layouts: `BorderLayout`, `BoxLayout`, `GridLayout`, and `FlowLayout`. Any one of these layouts can be applied towards the content of a `Panel` or `Frame` container. Although each layout in the library is represented with a different prototype, those are only templates initialized with basic layout parameters, such as vertical or horizontal gap. The actual arrangement of components is performed by the `LayoutManager` prototype, instantiated inside both `Panel` and `Frame` structures.

`LayoutManager` requires access to an instance of the preferred layout prototype and children nodes of the affected container. Whenever necessary, the container invokes the `LayoutManager`'s `doLayout` function, which forces all content to be rearranged according to the current profile. Based on the layout choice and dimensions of the parent, the children are traversed and individually wrapped into `Transform` holders with calculated horizontal and vertical offsets. At the end of `doLayout` operation `LayoutManager` also reports the ascertained minimum width and height of the container, which are used to update—if necessary—the actual size, and properly constrain the future resizes of the container.

*4.6.2 BorderLayout*
`BorderLayout` is employed to place subcomponents in up to five areas: `NORTH`, `SOUTH`, `WEST`, `EAST`, and `CENTER`. The unused space is allotted to the `CENTER` area. This is the only layout that requires the prior knowledge about the area that each child should be assigned to. When the `BorderLayout` partitions the container's surface into several regions, it first compares the available space with the cumulative size of the children's dimensions. If occlusions are expected, the container has to be enlarged; if, to the contrary, extra space is left, the peripheral components are pushed to the edges, while the center component is placed in the middle of the remaining space.

*4.6.3 BoxLayout and GridLayout*

With the use of the `BoxLayout` prototype, components in a `Panel` or `Frame` can be arranged in a single row or column, and additionally aligned vertically and horizontally. The 'Temperature Control' window from Figure 6 shows an example of using a `BoxLayout` with vertical orientation.

`GridLayout` allocates components to individual cells of a grid that contains the requested number of rows and columns. By using the `compressHorizontally` and `compressVertically` flags, the layout can be instructed to either level all rows in height and all columns in width, or condense each row and column on an individual basis. In situations, when the number of children does not equal the number of cells in the grid, vacant cells will be left empty, and redundant elements will not be rendered.

*4.6.4 FlowLayout*

`FlowLayout`, used by default in every `Panel` and `Frame` node, puts elements in rows, "jumping" to a new row every time the remaining horizontal space of the current row is insufficient for the next item. `FlowLayout` is the most sophisticated of all layouts implemented in X3DUI, because, depending on the current size of the container, the number of components in any given row as well as position of any given component in the row may vary. As a result, the minimum width and height that the container can be resized to on a unilateral operation might not agree with the permissible size for a bidimensional operation. In other words, resizing a window horizontally first and vertically second may produce a different result from applying the transformation in both directions simultaneously.

## 4.7 Deployment

In the development environment all twenty-seven prototypes of the library are stored in separate files, located in four directories. Another directory contains the graphical files. While the total size of graphics is less than 4 KB, and it is not feasible to try merging them into one file, the source code measures just under 430 KB and can be effectively integrated into a single resource. We have built an automated tool that generates a single file containing all prototypes and devoid of redundant spaces. The tool yields minified X3D code of about 280 KB, which equates to a 35% reduction. More importantly, the scene-loading time is accelerated by fewer file-system requests and prototype-scope sharing, resulting in elimination of external-prototype declarations.

However, when an X3D project includes the X3DUI library, a separate `ExternProtoDeclare` statement should be added for every prototype that will be used in the scene; the URL parameter must contain a relative or absolute path to the library file as well as the anchor to the referenced prototype. If the folder with images is removed from the original location within the library package, the new location has to be specified in the corresponding attribute of the `Display` prototype. When valid references are observed, X3DUI is readily deployable on the Web as a part of larger X3D visualization systems.

## 5. CONCLUSION

### 5.1 Summary

This paper has presented our work on the design and development of X3DUI, a GUI library for the X3D modeling language. We have reviewed and classified several existing approaches to building GUIs in X3D-driven visualization systems. We have also identified the main problems of current systems and addressed them natively in X3D by using special implementation techniques. The specificity of the language has been considered to ascertain a number of advisable usability-oriented practices employed in X3DUI. Finally, the organization of the library and essential characteristics of its nodes have been discussed.

### 5.2 Assessment

Although no actual assessment of the X3DUI library has been conducted yet, we are confident that our research and development efforts could be of interest to the Web3D community. Even in its early stages, X3DUI already demonstrates the qualities of a promising GUI framework. The major advantages of the library are

- *Adaptability*. X3DUI dynamically adjusts the GUI appearance to different resolutions and screen sizes, and can be configured with customized visual themes.
- *Compatibility*. Implemented entirely in X3D, lightweight, and easy-to-deploy, the library is integratable with many existing Web3D solutions, and suitable for stereographic imaging.
- *Efficiency*. Simple geometry, minimal use of texture graphics, and undemanding computation cycles make X3DUI suitable for both desktop and Web-based visualizations.
- *Flexibility*. Wide functional diversity of the library components allows X3D authors to better tailor the "look-and-feel" of the application to the project-specific needs.
- *Neatness*. Appealing exterior, smooth rendering, and natural blending of 2D and 3D ingredients enable X3DUI apt for quality production applications.
- *Operability*. Intuitive navigation across a variety of conventional UI components ensures intelligible and predictable interaction with the scene.
- *Reusability*. The same GUI prototypes can be reused to provide unified look across many applications.

The main limitation of X3DUI is the use of proprietary functions and nodes that are supported by one particular X3D player. Nonetheless, the functionality available through the Bitmanagement Software X3D extension answers the needs of numerous ongoing developments and appears relevant to the modern Web3D design trends. Hence we believe that the majority of nonstandard elements employed in X3DUI should become a part of the X3D specification. For instance, `Layer3D` and `OrderedGroup` nodes are irreplaceable for multi-scene management and dynamic z-depth stack control. The advantages of texture-based text rendering, achievable with `USE_TEXTURE` flag, were discussed in section 3.3. The ability to request the type, name, and bounding dimensions of an object (accomplished with

`getType`, `getName`, and `getBBox` functions, correspondingly) can lead to significant improvements of scripting and rendering performance. Finally, the `Browser` object extensions allowing to monitor the current size of the screen (`windowSize` attribute with `getWindowSizeX` and `getWindowSizeY` functions); control visibility of the cursor (`hideCursor` function); set the navigation mode (`setNavigationMode` function); and create new nodes at runtime (`createVrmlFromString` function)—undoubtedly provide better integration with the host OS and enhance to the dynamics of the scene.

### 5.3 Future Work

The development of X3DUI is still in progress and requires substantial revision before a fully functional release is produced. One of the priority directions for future work is the reduction of recurring code patterns and further delegation of subsidiary tasks to designated prototypes. Another important goal is altering the programming logic to recognize different X3D player engines and serve only the player-supported content.

With regards to the component diversity, the library could be expanded with several unimplemented GUI widgets including text area control, toolbars, file and context menus, dialogs, vertical and horizontal scrolls, lists, icons, and so on. The existing nodes could be supplemented with additional functionality as well. For example, the overall interface accessibility could be improved with tabbing navigation and shortcut support.

To minimize the development-to-production overhead, we have built a software tool to automatically generate user documentation from source code—similarly to Javadoc [14]; attribute- and method-specific descriptions are supplied within XML comment tags.

An intriguing idea, inspired by the success of Google Web Toolkit [15], is to build an X3DUI development suite that would translate GUIs written in a modern object-oriented language, such as Java, into X3D code. Due to the high level of abstraction, the programmer would not require the extensive expertise in X3D.

## 6. ACKNOWLEDGMENTS

Our thanks to BitManagement Software GmbH for providing us with their Development Toolkit.

## 7. REFERENCES

[1] Hamza-Lup, F. G., Sopin, I., and Zeidan, O. 2007. Online towards 3D Web-based simulation and training systems for radiation oncology. ADVANCE Magazine for Imaging and Oncology Administrators. 17, 7 (July 2007). 64–68.

[2] Dachselt, R., Hinz, M., and Meissner, K. 2002. CONTIGRA: an XML-based architecture for component-oriented 3D applications. In Proceedings of the Seventh International Conference on 3D Web Technology (Tempe, AZ, February 24–28, 2002). ACM Press, New York, NY, 155–163.

[3] Taewoo, K. and Fishwick, P. 2002. A 3D XML-based customized framework for dynamic models. In Proceedings of the Seventh International Conference on 3D Web Technology (Tempe, AZ, February 24–28, 2002). ACM Press, New York, NY, 103–109.

[4] Calleros, J. M. G., Vanderdonckt, J., and Arteaga, J. M. 2006. A method for developing 3D user interfaces of information systems. In Proceedings of the Sixth International Conference on Computer-Aided Design of User Interfaces (Bucharest, Romania, June 6–8, 2006). Springer Netherlands, Dordrecht, the Netherlands, 85–100.

[5] Limbourg, Q., Vanderdonckt, J., Michotte, B., Bouillon, L., Florins, M., and Trevisan, D. 2004. UsiXML: a user interface description language for context-sensitive user interfaces. In Proceedings of the Ninth IFIP Working Conference on Engineering for Human-Computer Interaction (Hamburg, Germany, July 11–13, 2004). Springer-Verlag, Berlin, Germany, 200–220.

[6] Network-Enabled WorkSpaces. http://news.felixlup.info. June 2010.

[7] Souza, E., Roque, M., Silva, L., Cardoso, A., and Lamounier, E. 2007. Ludos Top: a Web-based 3D educational game multi-user online. http://seminfo.com.br/anais/2007/pdfs/11-34927.pdf. April 2010. Federal University of Uberlândia, Uberlândia, Brazil.

[8] Francis, B. and Stone, R. 2009. WebScylla: a 3D Web application to visualise the colonisation of an artificial reef. In Proceedings of the Fourteenth International Conference on 3D Web Technology (Darmstadt, Germany, June 16–17, 2009). ACM Press, New York, NY, 167–175.

[9] Behr, J., Eschler, P., Jung, E., and Zollner, M. 2009. X3DOM: a DOM-based HTML5/X3D integration model. In Proceedings of the Fourteenth International Conference on 3D Web Technology (Darmstadt, Germany, June 16–17, 2009). ACM Press, New York, NY, 127–135.

[10] Roberts, J., Yoon, I., Yoon, S., and Lank, E. 2004. An interface mark-up language for Web3D. In Proceedings of the Sixth IASTED International Conference on Software Engineering and Applications (Cambridge, MA, November 9–11, 2004). ACTA Press, Calgary, Canada, 518–523.

[11] Holmberg, N., Wunsche, B., and Tempero, E. 2006. A framework for interactive Web-based visualization. In Proceedings of the Seventh Australasian User Interface Conference (Hobart, Australia, January 16–19, 2006). Australian Computer Society, Inc., Sydney, Australia, 137–144.

[12] Cockburn, A. and McKenzie, B. 2002. Evaluating the effectiveness of spatial memory in 2D and 3D physical and virtual environments. In Proceedings of the SIGCHI Conference on Human Factors in Computing Systems: Changing Our World, Changing Ourselves (Minneapolis, MN, April 20–25, 2002). ACM, New York, NY, 203–210.

[13] Lindeman, R., Sibert, J., and Templeman, J. 2001. The effect of 3D widget representation and simulated surface constraints on interaction in virtual environments. In Proceedings of the Virtual Reality Conference (Yokohama, Japan, March 13–17, 2001). IEEE Computer Society, Washington, DC, 141–148.

[14] Javadoc Tool. http://java.sun.com/j2se/javadoc. June 2010.

[15] Google Web Toolkit. http://code.google.com/webtoolkit. June 2010.